# Assessment of Three Databases for the NASA Seven-Coefficient Polynomial Fits for Calculating Thermodynamic Properties of Individual Species

Research Article

Osama A. Marzouk*

Assistant Professor, College of Engineering, University of Buraimi, Al Buraimi, Sultanate of Oman, Oman.

**Abstract**

This work considers the seven-coefficient polynomials proposed by the National Aeronautics and Space Administration (NASA) to facilitate obtaining a normalized value for three thermodynamic standard-state specific properties of ideal gases or condenser matters over an interval of temperature. These properties are the heat capacity at constant pressure, the absolute enthalpy (sensible enthalpy plus heat contents due to chemical or physical changes), and the entropy. In the open literature, one can find several databases for the polynomial coefficients with variation in the number of species included or the range of temperature covered, and this raises a question of whether the choice of a database to use has an important impact on these evaluated thermodynamic properties. Addressing this point, we compare and assess three databases for the NASA 7-coefficient polynomials, over a selected range of temperature from 300 K to 3500 K, and for selected six common gaseous species encountered in combustion or industrial processes, which are molecular oxygen ($O_2$), molecular nitrogen ($N_2$), molecular hydrogen ($H_2$), methane ($CH_4$), carbon dioxide ($CO_2$), and water vapor ($H_2O$). Our comparisons suggest that despite the difference in the values of coefficients, there is no significant difference in their predictions. However, the latest ($7^{th}$ edition) database of Prof. Alexander Burcat hosted at Eötvös Loránd University (ELTE) in Budapest, Hungary showed superior features when contrasted to other two databases (one accompanying the simulation package OpenFOAM 6, and another provided by the natural-gas reaction mechanism GRI-MECH 3.0).

**Abbreviations:** NASA: National Aeronautics and Space Administration; ELTE: Eötvös Loránd University; GRI: Gas Research Institute; IGT: Institute of Gas Technology; GTI: Gas Technology Institute; CFD: Computational Fluid Dynamics; NIST: National Institute of Standards and Technology; GRI-MECH: Gas Reaction Mechanism.

## Introduction

There is a number of analytical fittings for thermodynamic properties of ideal gases or condensed matters, based on data collected experimentally or using results of statistical calculations. These expressions have the temperature as their independent variable. Having an analytical function for these properties is a large advantage in computerized modeling, as compared to tables, especially when the evaluation of the properties has to performed many times before reaching a final solution. One of the early efforts in this area was made by Shomate [1]. The National Aeronautics and Space Administration (NASA) has also developed a set of three polynomial functions for standard-state specific heat at constant pressure, standard-state specific enthalpy, and standard-state specific entropy for use in a program code that performs chemical equilibrium calculations for a variety of settings [2]. This set contains 7 coefficients for each interval of temperature for each species, and they are obtained using a least-squares method [3] to match the polynomials predictions to a reference set of data. Each species can have more than one temperature interval, and each interval will have its own set of 7 coefficients. We refer to this set of polynomials as the NASA seven-coefficient polynomial fits, or simply NASA polynomials. The early database of coefficients published by NASA covered 224 species and two temperature intervals for most of these substances: from 300 K to 1000 K, and from 1000 K to 5000 K. This database has been updated continually until 1993 [4].

Later, NASA has adopted a more-detailed set of polynomials by adding two additional terms and coefficients for each of the three functions, leading to a new 9-coefficient version of the NASA polynomials. The updated 9-coefficient polynomials are accompanied with a more-comprehensive database for coefficients that expand from 200 K to 20000 K. This wide range can be of interest for particular space applications, but for terrestrial combustion and industrial process as well as chemical research and energy systems, such high temperatures are not normally expected. Despite this, the format of the 7-coefficient polynomials has been accepted widely by computational modelers and incorporated in a number of simulation software packages [5-11]. In this work, the term "NASA polynomials" refers to the older version with 7 coefficients.

To use the NASA polynomials, one needs a database for the coefficients. There are different freely-available databases in the literature, with varying number of involved species and temperature ranges. We here consider three databases for the NASA coefficients, which are:

### I. GRI-MECH 3.0 [12]

This database is housed at the University of California at Berkeley. It is a part of experimental and computations work performed at the University of California at Berkeley, Stanford University, University of Texas at Austin, and SRI International (a nonprofit research center in the USA, formerly Stanford Research Institute [13]). This work is known as the GRI-Mech Project, which is a collection of chemical reactions and associated expressions for the rate constants, for modeling the combustion of natural gas. The work was sponsored by the Gas Research Institute (GRI) as a major research and technology organization serving the natural gas industry at that time, which later combined with the Institute of Gas Technology (IGT) to form the Gas Technology Institute (GTI) in April 2000 [14].

The number of species in this database is 53. The database we consider here is part of version 3.0 of this project, which is the latest version at the time of preparing this manuscript. The database is downloadable as a fixed-format text file at http://combustion.berkeley.edu/gri-mech/version30/files30/thermo30.dat.

### II. OpenFOAM 6 [15]

OpenFOAM is an open-source computational fluid dynamics (CFD) solver that can handle a wide variety of fluid, thermal, and combustion problems. This package gives the user a number of options for modeling the thermodynamic properties, and one of them is called (janaf). It is basically the NASA polynomials but with coefficients based on tables of JANAF (Joint Army-Navy-Air Force)[16, 17]. The latest edition of JANAF tables published in paper is the 4th edition [18]. JANAF tables are now accessible online at the National Institute of Standards and Technology (NIST) [19]. It is worth mentioning that the first edition of the JANAF tables (1960) was not publicly published, but was internally available to agencies of the United States government [20].

The number of species in this database is 1489. The database we consider here accompanies version 6 of OpenFOAM files, which is the latest version at the time of preparing this manuscript. The software source files are downloadable as a tar.gz compressed archive file at http://dl.openfoam.org/source/6. This needs to be decompressed and after extracting the contained files as a folder tree; one can find a number of database test files in a fixed-format style. We used the largest file in the tutorials, which is located at the local path: OpenFOAM-6-version-6\tutorials\combustion\chemFoam\ic8h18_TDAC\chemkin\therm.dat, where (OpenFOAM-6-version-6) is the root folder made after extracting files from the archive file.

### III. Burcat 7th [21]

The last database we analyze here is the latest edition provided by Prof. Alexander Burcat for thermodynamic data, located at the web site of the Laboratory for Chemical Kinetics, Institute of Chemistry at Eötvös Loránd University (ELTE), Budapest, Hungary. It is mirrored daily from the original Technion web site: http://burcat.technion.ac.il/dir.

The number of species in this database is 2972. The database we consider here is the 7th edition, which is the latest version at the time of preparing this manuscript. The database is downloadable as a fixed-format text file at http://garfield.chem.elte.hu/Burcat/THERM.DAT.

## Mathematical Relations

We start with the functional form for the standard-state specific heat capacity at constant pressure, which is a fourth-order polynomial in temperature as given in Eq. (1).

$$\overline{\frac{C_p^\circ}{R}} \text{ or } \frac{c_p^\circ}{R} = a_1 + a_2 T + a_3 T^2 + a_4 T^3 + a_5 T^4 \quad \text{----(1)}$$

where the coefficients ($a_1-a_5$) are subset of the full set of 7 coefficients (per temperature interval, per species), and $T$ is the absolute temperature (in kelvins).

The superscript ° indicates a standard state, which is conventionally either 1 bar ($10^5$ Pa) or 1 atm (101,325 Pa). JANAF tables used to adopt 1 atm in its 1st and 2nd editions, but in the 3rd and 4th editions, the 1-bar reference was adopted for all species. This is also followed by NASA in its extended 9-coeffecient database (but for gases only, whereas for condensed species the standard pressure is 1 atm) [22]. In our work, all species we consider are gaseous, leading to an agreement on the standard-state pressure to be 1 bar by both the NASA extended database and the JANAF tables.

The effect of standard-state pressure difference (between 1 atm and 1 bar) is a small change in some thermodynamic quantities for the species. For condensed phases, the magnitude of these changes is nearly always negligible relative to the uncertainty of data. For gaseous species, the changes in the standard-state entropy may not be negligible [18]. For perfect (ideal) gas, the specific heat capacity and specific enthalpy (and specific internal energy) depend on the temperature only, and the standard-state values are the actual values [5, 23]. With this approximation, the matter of standard-state pressure value becomes irrelevant. Reference [18] has formulas that perform corrections for the standard-state entropy in case of non-ideal gaseous species (no correction is considered important for the enthalpy and specific heat capacity or for condensed-phase species).

Equation (1) is in a normalized form, so it gives a nondimensional value. Multiplying the obtained nondimensional value by the universal gas constant ($\overline{R}$) gives the molar standard-state specific heat capacity ($\overline{C_p^\circ}$) in the same units of the utilized universal gas constant. One may multiply the nondimensional value by the specific gas constant ($R$) to obtain the mass-based standard-state specific heat capacity ($c_p^\circ$), where

$$R = \frac{\overline{R}}{M} \quad \text{-----(2)}$$

and ($M$) is the molecular weight. Alternatively, one can obtain the mass-based value as

$$c_p^\circ = \frac{\overline{C_p^\circ}}{M} \quad \text{-----(3)}$$

The standard-state molar specific enthalpy ($\overline{H^\circ}$) is the integral of the standard state molar constant-pressure specific heat capacity with respect to the absolute temperature. The standard-state mass-based specific enthalpy ($h^\circ$) can be obtained in a similar integral using the standard-state mass-based constant-pressure specific heat capacity. Thus,

$$\overline{H^\circ} = \int \overline{c_p^\circ} dT; \quad h^\circ = \int c_p^\circ dT \quad \text{-----(4)}$$

Applying this to the polynomial of the specific heat capacity in Eq. (1), we obtain a second polynomial for the standard-state molar enthalpy. To make it nondimensional, we divide both sides by the temperature to obtain

$$\frac{\overline{H^\circ}}{\overline{R}T} \text{ or } \frac{h^\circ}{RT} = a_1 + \frac{a_2}{2}T + \frac{a_3}{3}T^2 + \frac{a_4}{4}T^3 + \frac{a_5}{5}T^4 + \frac{a_6}{T} \quad \text{-----(5)}$$

where ($a_6$) is an enthalpy integration constant. One can also obtain ($h^\circ$) from $\overline{H^\circ}$ using

$$h^\circ = \frac{\overline{H^\circ}}{M} \quad \text{-----6}$$

Scientifically, the molar specific enthalpy for an individual species is defined as

$$\overline{H^\circ}(T) \equiv \overline{H^\circ}(T_{ref}) + [\overline{H^\circ}(T) - \overline{H^\circ}(T_{ref})] = \overline{H^\circ}(T_{ref}) + \int_{T_{ref}}^{T} \overline{C_p^\circ}(T) \, dT \quad \text{----7}$$

where ($T_{ref}$) is an arbitrary reference temperature (298.15 K is a common value). This accounts for both the sensible component and any heat gain due to physical or chemical change (e.g., heat of vaporization). However, the calculated specific enthalpy by the polynomial has a slightly different definition where the specific standard-state enthalpy of formation at the reference temperature $\overline{\Delta H_f^\circ}(T_{ref})$ is used instead of the specific standard-state enthalpy itself at the reference temperature $\overline{H^\circ}(T_{ref})$. Thus,

$$\overline{H^\circ}(T) = \overline{\Delta H^\circ}(T_{ref}) + [\overline{H^\circ}(T) - \overline{H^\circ}(T_{ref})] = \overline{H_f^\circ}(T_{ref}) + \int_{T_{ref}}^{T} \overline{C_p^\circ}(T) \, dT \quad \text{-----8}$$

We call this form of enthalpy (engineering absolute enthalpy), since this format is common in engineering [20]. Equations (7) and (8) coincide if one takes the enthalpy of formation at the reference temperature as the reference enthalpy for each species,

$$\overline{H^\circ}(T_{ref}) \equiv \overline{\Delta H_f^\circ}(T_{ref}) \quad \text{-----9}$$

For a single species, the datum enthalpy value is completely arbitrary, and Eq. (9) is always an acceptable choice for setting a datum enthalpy.

We point out that the integration constant ($a_6$) is not necessarily equal to $\overline{\Delta H_f^\circ}(T_{ref})/\overline{R}$.

The standard-state specific enthalpy (either molar $\overline{s^\circ}$, or mass-based $s^\circ$) and standard-state specific heat capacity at constant pressure are related by

$$\overline{S^\circ} = \int \frac{\overline{C_p^\circ}}{T} dT; \quad s^\circ = \int \frac{c_p^\circ}{T} dT \quad \text{-----10}$$

From Eq. (1), this gives the following expression for the normalized standard state entropy

$$\frac{\overline{S^\circ}}{\overline{R}} \text{ or } \frac{s^\circ}{R} = a_1 \ln T + a_2 T + \frac{a_3}{2}T^2 + \frac{a_4}{3}T^3 + \frac{a_5}{4}T^4 + a_7 \quad \text{-----11}$$

where ($a_7$) is an entropy integration constant. One can also obtain the mass based ($s^\circ$) from the molar ($\overline{S^\circ}$) using

$$s^\circ = \frac{\overline{S^\circ}}{M} \quad \text{-----12}$$

In some open-system devices with a gas flow, like nozzles and diffusers, the flow may be approximated as isentropic (no change in specific entropy) [24]. Also, in gas dynamics the isentropic condition is commonly used for the sake of a simplified analysis with closed-form solutions or reduced governing equations [25, 26]. In such cases, an important gas property needed for the analysis appears, which is the specific heat ratio (or adiabatic index), γ.

$$\gamma = \frac{c_p}{c_v} = \frac{\overline{C_p}}{\overline{C_v}} \quad \text{-----13}$$

where ($c_v$) and ($\overline{C_v}$) are the mass-based and molar specific heat capacities at constant volume, respectively. They are related as

$$c_v = \frac{\overline{C_v}}{M} \quad \text{-----14}$$

The constant-pressure and constant-volume specific heat capacities are related as follows:

$$\overline{C_v} = \overline{C_p} - \overline{R} \quad \text{----- 15}$$

$$c_v = c_p - R \quad \text{-----16}$$

Under the assumption of ideal gas, the standard-state values of the specific heat at constant pressure become the actual values regardless of the pressure. In this case, one can use these standard-state values as obtained using the polynomials to calculate the actual specific heat ratio. This gas property is the fourth item we look at when comparing the databases of polynomial coefficients. We add that the ideal-gas assumption is a good approximation at normal ambient conditions, where most real gases behave qualitatively as ideal gases [27]. Water vapor can be treated as an ideal gas only when not near the saturated-vapor state, such as moisture in air conditioning [28] or high superheat steam in steam turbines [29].

## Species and Temperatures Under Study

It is not practical to compare or assess the database of polynomial coefficients for all species they cover, because some databases cover a large number of species and considering all of them is a huge amount of work, and also because different databases cover different numbers of species and thus a species my not exist in all databases being analyzed. We consider 6 species in this work that are of special importance in combustion applications or energy systems, and also commonly included in all the databases we examine here.

These species are:

1. Molecular oxygen ($O_2$)
2. Molecular nitrogen ($N_2$)
3. Molecular hydrogen ($H_2$)
4. Methane ($CH_4$)
5. Carbon dioxide ($CO_2$)
6. Water vapor ($H_2O$)

Oxygen and nitrogen are the major components of air, so classified as oxidizers. Hydrogen and methane are fuels. Carbon dioxide and water vapor are combustion products.

Regarding the temperatures to be examined here, they are the same for all the six species under study, which is from 300 K to 3500 K. This range ensures consistency in comparison because it is common across all three databases for the six gaseous species. We do not examine the full temperature range of each individual species at each databases; this varies from one database to another.

All the three databases have two temperature intervals for the six species: low-temperature interval, and high-temperature interval. The common temperature is where the two intervals meet.

The temperatures (in kelvins) at which the polynomials were evaluated are: 300, 500, 750, 1000, 1250, 1500, 1750, 2000, 2250, 2500, 2750, 3000, 3250, and 3500. Thus, the temperature range we study is discretized into 14 points (13 steps).

## Constants and Reference Conditions

A number of constants or fixed values are used throughout this study in the calculation. Although some constants are universal, the exact value may vary slightly from one source to another. We here give the values utilized in the current work, starting with the universal gas constant ($\bar{R}$) and the reference temperature ($T_{ref}$) and standard-state pressure ($p°$) in Table 1.

Table 2 gives the molecular weights for the studied species, and the corresponding value of the specific gas constant as calculated using Eq. (2).

The polynomial coefficients themselves are listed along with the intervals of temperature in Table 11 for $O_2$, Table 12 for $N_2$, Table 13 for $H_2$, Table 14 for $CH_4$, Table 15 for $CO_2$, and Table 16 for $H_2O$. These tables are given in **Appendix A**.

## Results

### Number of Species

In terms of the number of species covered by the database, the Burcat database comes in the first place among the three databases we analyze, with a total of 2972 species.

### Range of Temperature

In terms of the range of temperatures covered for the six gaseous species we consider here, the Burcat database comes in the first place among the three databases we analyze, as it has the widest

**Table 1. Values of the universal gas constant and reference values used in the calculations.**

| Universal gas constant ($\bar{R}$) | Reference temperature for the enthalpy ($T_{ref}$) | Standard-state pressure ($p°$) |
|---|---|---|
| 8.31451 J/mol-K [22] | 298.15 K [18, 22] | 1 bar [18, 22] |

**Table 2. Molecular weights and specific gas constants of considered species.**

| Species | Molecular weight, M (g/mol) [22] | Specific gas constant, R (J/kg-K) |
|---|---|---|
| Molecular oxygen ($O_2$) | 31.9988 | 259.838 |
| Molecular nitrogen ($N_2$) | 28.0134 | 296.805 |
| Molecular hydrogen ($H_2$) | 2.01588 | 4124.51 |
| Methane ($CH_4$) | 16.04246 | 518.281 |
| Carbon dioxide ($CO_2$) | 44.0095 | 188.925 |
| Water vapor ($H_2O$) | 18.01528 | 461.525 |

range 0f 200 K to 6000 K for each of these species.

**Continuity Test**

We evaluated the four standard-state gas properties using each database at the common temperature point (where the low-temperature interval and the high-temperature interval meet). Since the common temperature belongs to both intervals, the coefficients of either of the two intervals should give the same result. We found that this is practically satisfied by all databases. The OpenFOAM showed a one-digit difference for the estimated standard-state mass-based specific enthalpy of $O_2$, $H_2$, and $H_2O$. However, this is practically null compared to the large values of that property.

**Enthalpy of Formation**

The ability of the analyzed databases to generate a proper value for the enthalpy of formation at the reference temperature (298.15 K) is tested here. As in Eq. (9), this value should be the estimated enthalpy by the enthalpy polynomial at this temperature. Table 3 lists the estimations from the databases along with a benchmarking value reported in the JANAF tables (latest published edition). It should be noted that the low-temperature coefficients (not the high-temperature ones) were used in the estimations since their interval includes this reference temperature.

Unless the JANAF value is zero or the reference temperature is outside the range of the database, the percentage deviation is also reported in the table, which we define as

$$\% \text{ deviation} = \frac{(\text{database value} - \text{JANAF value})}{\text{JANAF value}} \times 100\% \quad \text{-----17}$$

The GRI interval for $N_2$ ends at 300 K at its lower bound, which is slightly above the reference temperature. The OpenFOAM database gives a non-zero value for all the diatomic species, whereas the JANAF value is zero. For $CH_4$, $CO_2$, and $H_2O$, the magnitudes of the calculated deviations are very small or even zero practically, with the worst case being (0.365%) for $CH_4$ when using either the GRI database or the Burcat database.

**Extrapolation to Absolute-Zero Temperature**

According to Eq. (5), at T = 0 K, the estimated value of $\frac{\overline{H}^°}{R}$ or $\frac{h^°}{R}$ is ($a_6$), and this has a unit of K. There are two values of ($a_6$) in each of the analyzed databases here: a low-temperature value and a high temperature value. The low-temperature ($a_6$) is considered here as it belongs to the interval closer to the 0 K. As a benchmarking value, we also list the value estimated from a Technical Paper by NASA [22].

A summary of the results is given in Table 4. We point out that this part should be viewed carefully keeping in mind that all the databases are used outside their lowest temperature value. Thus, it is not necessary that they will give a value that agrees well with the benchmarking one. One can view this part as subsidiary information, rather than a test of goodness. Despite this note of caution, all databases show rough agreement with the benchmarking values.

**Variation with Temperature**

Visual comparisons of the profile of the four standard-state gas properties (specific heat at constant pressure, specific heat ratio, specific enthalpy, and specific entropy) are presented graphically here. The specific quantities are mass based (per kilogram). For $N_2$, the GRI and the OpenFOAM databases have almost the same polynomial coefficients, leading to nearly identical estimates.

Figure 1 shows the temperature profiles for the standard-state mass-based specific heat capacity at constant pressure for the two oxidizer species: $O_2$ and $N_2$. The estimates from the three databases are not identical, but are very close to each other. For both species, the specific heat increases with temperature but with different patterns, where $N_2$ tends to saturate earlier than $O_2$.

The temperature profiles for the standard-state mass-based specific heat capacity at constant pressure for the two fuel species: $H_2$ and $CH_4$ are given in Figure 2. There is a good agreement among the estimates from the three databases in the case of $H_2$ (but the estimates are not identical). In the case of $CH_4$, the GRI database deviates at a temperatures of 1750 K and above from the other two databases and gives higher values, and also shows slower saturation behavior. At a temperature of 3500 K (highest in the range studied here), the value estimated by the GRI database is 7,188.5 J/kg-K. On the other hand, the OpenFOAM database gives 6,270.5 J/kg-K, and the Burcat database gives 6,415.3 J/kg-K. As a benchmarking value, we use the value listed in the last (4th) printed edition of the JANAF tables [18], which is 103.060

Table 3. Comparison of the mass-based specific enthalpy (in J/kg) at $T_{ref}$ = 298.15 K (and percentage deviation if defined) as predicted by the polynomial coefficients for the low-temperature interval and the one reported in the JANAF tables (4th edition [18]) for the considered species.

| Species | GRI | OpenFOAM | Burcat | JANAF |
|---|---|---|---|---|
| Molecular oxygen ($O_2$) | 0 | -26 | 0 | 0 |
| Molecular nitrogen ($N_2$) | Out of range | 51 | 0 | 0 |
| Molecular hydrogen ($H_2$) | 0 | 1,214 | 0 | 0 |
| Methane ($CH_4$) | -4,650,160 (-0.365%) | -4,666,359 (-0.0175%) | -4,650,160 (-0.365%) | -4,667,177 |
| Carbon dioxide ($CO_2$) | -8,941,479 (-0.00305%) | -8,942,340 (0.00658%) | -8,941,479 (-0.00305%) | -8,941,751 |
| Water vapor ($H_2O$) | -13,423,383 0% | -13,424,587 (0.00897%) | -13,423,383 0% | -13,423,383 |

Table 4. Comparison of the ($a_6$) polynomial coefficient for the low-temperature interval with a benchmarking value, which is the estimated zero-kelvin molar enthalpy (expressed as a temperature in K) of the considered species.

| Species | GRI | OpenFOAM | Burcat | $°(0)$ |
|---|---|---|---|---|
| Molecular oxygen ($O_2$) | -1.06394356E + 03 | -1.00524900E + 03 | -1.06394356E + 03 | -1.044E + 03 |
| Molecular nitrogen ($N_2$) | -1.02089990E + 03 | -1.02090000E + 03 | -1.04697628E + 03 | -1.043E + 03 |
| Molecular hydrogen ($H_2$) | -9.17935173E + 02 | -1.01252100E + 03 | -9.17935173E + 02 | -1.018E + 03 |
| Methane ($CH_4$) | -1.02466476E + 04 | -1.01424099E + 04 | -1.02453222E + 04 | -1.018E + 04 |
| Carbon dioxide ($CO_2$) | -4.83719697E + 04 | 4.84162830E + 04 | -4.83719710E + 04 | -4.845E + 04 |
| Water vapor ($H_2O$) | -3.02937267E + 04 | -3.02081100E + 04 | -3.02937260E + 04 | -3.028E + 04 |

Figure 1. Standard-state mass-based constant-pressure specific heat capacity of molecular oxygen and molecular nitrogen versus temperature, calculated using three different databases.

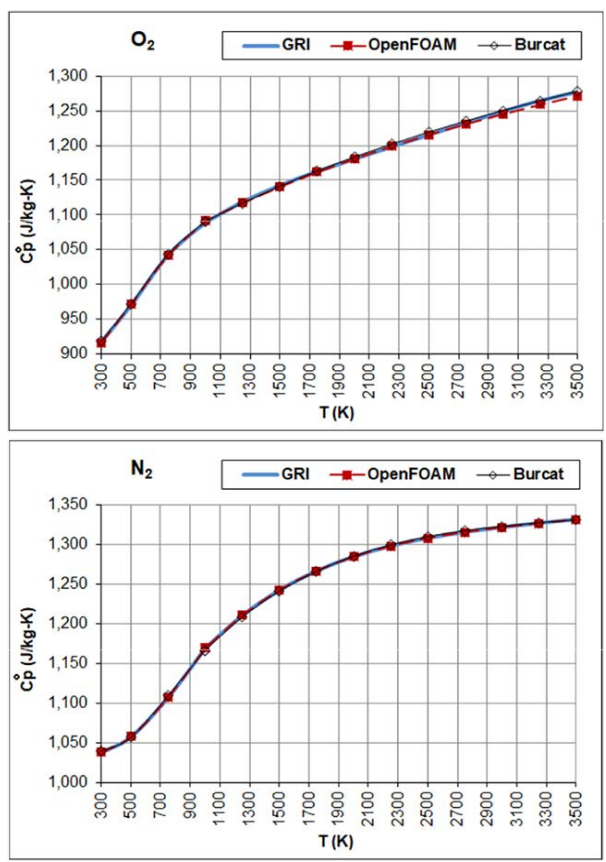

J/mol-K, corresponding to 6,457.7 J/kg-K (using a molecular weight of 16.04276 g/mol in the same edition). This value is closest to the value obtained using the Burcat database.

The temperature profiles for the standard-state mass-based specific heat capacity at constant pressure for the two combustion-products species: $CO_2$ and $H_2O$ are given in Figure 3. The estimates from the databases are not identical, but the deviations are small.

The temperature profiles for the standard-state specific heat ratio ($\gamma°$) for the six gaseous species under study are given in Figure. 4 for the oxidizer species, in Figure. 5 for the fuel species, and in Figure. 6 for the combustion-product species. Whereas the standard-state constant-pressure specific heat capacity was increasing with temperature, the standard-state specific heat ratio decreases with temperature, as expected from the relation

$$\gamma° = \frac{c_p°}{c_p° - R} \quad \text{-----18}$$

Both properties show monotonic nonlinear behavior in response to temperature change, with a decelerated change at high temperatures. The specific heat ratio appears in the exponent of some quantities in isentropic-flow analysis, which may make its impact significant. Polyatomic gases have a lower value than diatomic gases. Even diatomic gases show different impact of temperatures on their standard-state specific heat ratio (which is the actual one under the ideal-gas assumption). These points call for attention when attempting to assign a constant specific heat ratio in the analysis of a flow involving large variation in temperatures and/or compositions, such as flames, internal combustion engines, and shock waves, as followed in some books [30-32] or educational resources [33, 34]. As a quantitative example, the estimated standard-state specific heat ratio for $O_2$ at 300K is 1.395 (GRI and Burcat databases) and 1.396 (OpenFOAM database), whereas the corresponding value for $CO_2$ at 3500K is 1.153 (GRI and Burcat databases) and 1.156 (OpenFOAM database). Relative to the $O_2$ values, the $CO_2$ values are dropped by about 17.2%. This would be lost if one assumes a constant value for the entire domain.

Figure 2. Standard-state mass-based constant-pressure specific heat capacity of molecular hydrogen and methane versus temperature, calculated using three different databases.

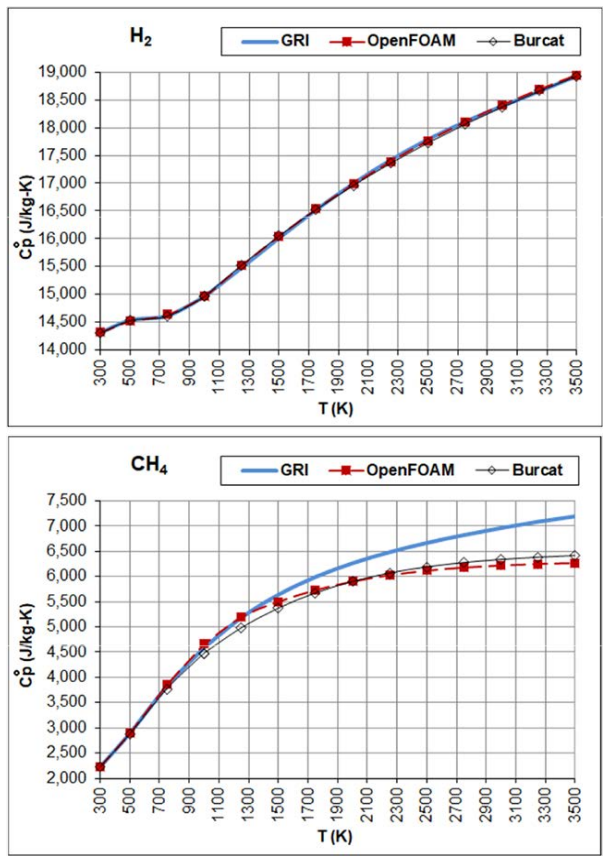

Figure 3. Standard-state mass-based constant-pressure specific heat capacity of carbon dioxide and water vapor versus temperature, calculated using three different databases.

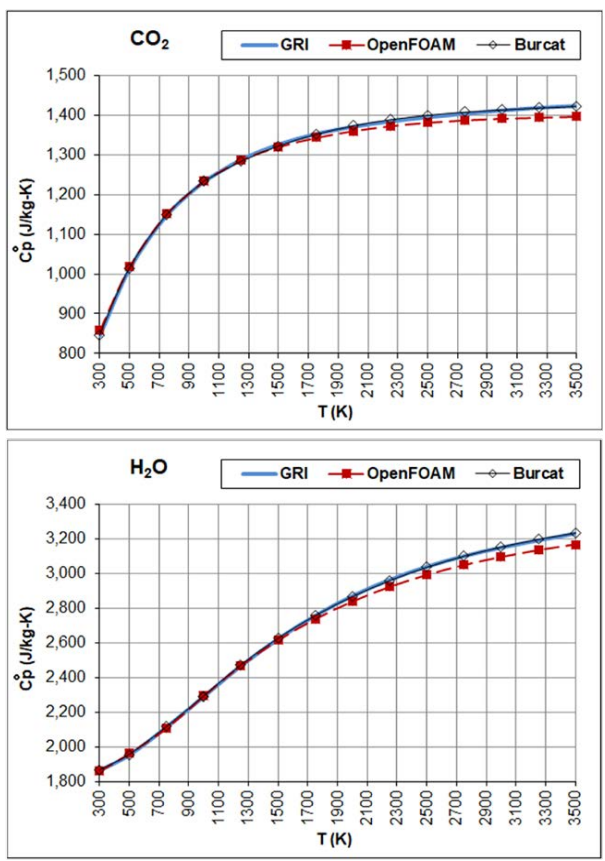

Figure 4. Standard-state specific heat ratio of molecular oxygen and molecular nitrogen versus temperature, calculated using three different databases.

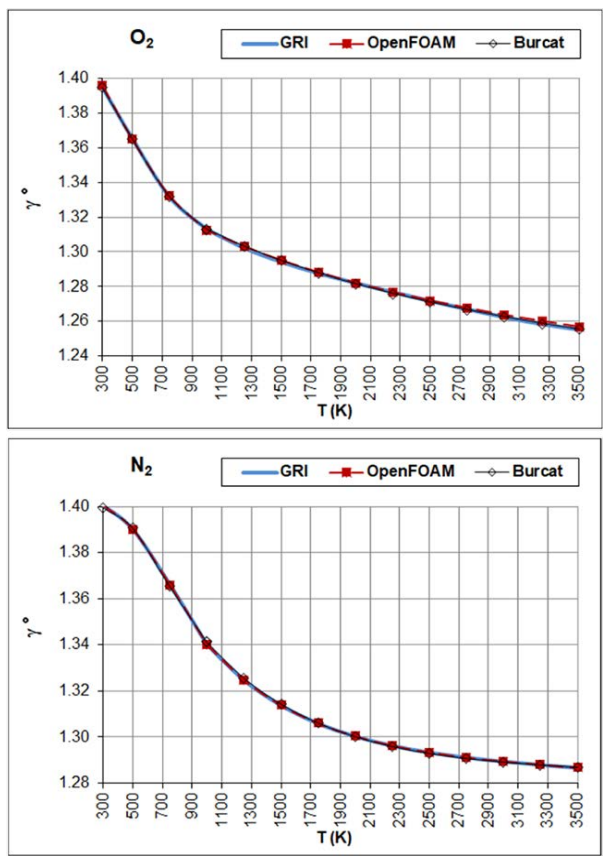

Figure 5. Standard-state specific heat ratio of molecular hydrogen and methane versus temperature, calculated using three different databases.

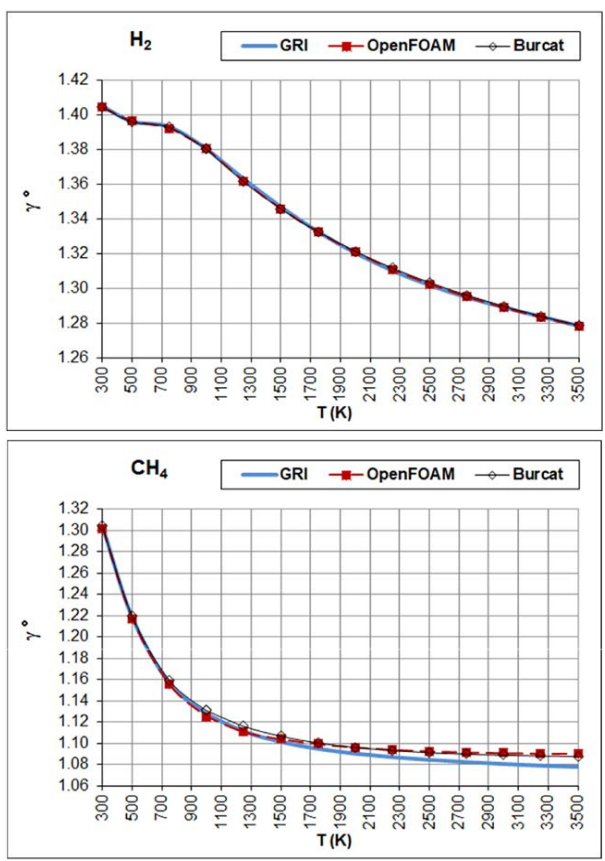

**Figure 6. Standard-state specific heat ratio of carbon dioxide and water vapor versus temperature, calculated using three different databases.**

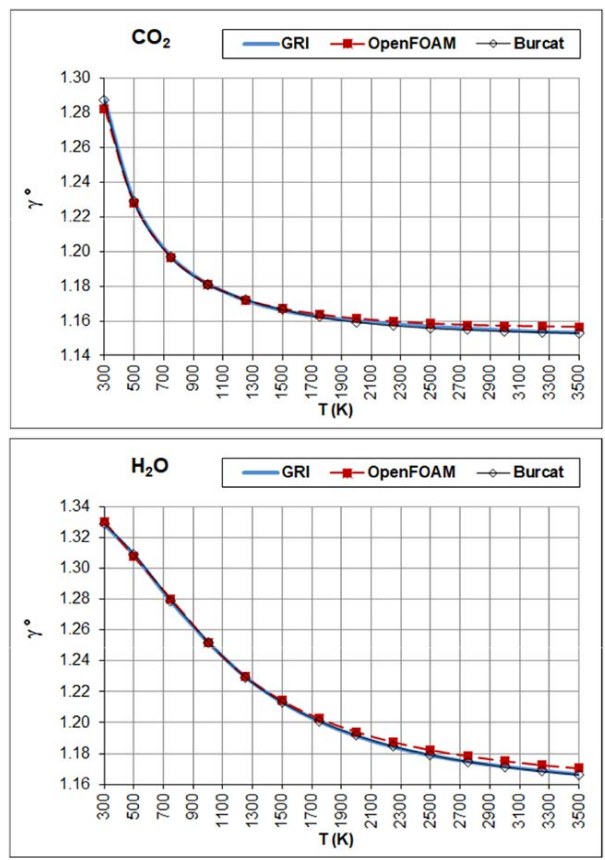

The profiles shown for the standard-state specific heat ratio show only small deviations due to adopting different databases of polynomial coefficients. Even the case of methane that showed remarkable deviation of the GRI database does not show this feature now. This can be explained by the form of Eq. (18), where the standard-state specific heat capacity appears in both the numerator and the denominator, thereby relieving its influence.

The temperature profiles for the standard-state mass-based absolute engineering specific enthalpy for the six gaseous species under study are given in Figure 7 for the oxidizer species, in Figure 8 for the fuel species, and in Figure 9 for the combustion-product species.

This thermodynamic property is based on the integration of the standard-state specific heat capacity at constant pressure. Thus, it shows less variability with regard to temperature than the standard-state specific heat capacity. The pattern of the profiles qualitatively resembles a linear function, and the deviations among the three databases are relatively not significant, with the exception of $CH_4$.

The molecular diatomic species ($O_2$, $N_2$, and $H_2$) have zero or positive values, whereas as the product species ($CO_2$ and $H_2O$) have negative values. This is a result of the datum specific enthalpy of formation at $T_{ref}$ = 298.15 K for these species, which is either zero in the case of $O_2$, $N_2$, and $H_2$, or a large negative value in the case of $CO_2$ and $H_2O$.

The temperature profiles for the standard-state mass-based specific entropy for the six gaseous species under study are given in Figure 10 for the oxidizer species, in Figure 11 for the fuel species, and in Figure 12 for the combustion-product species. This property increases nonlinearly and monotonically with the temperature, with a decelerated rate at high temperatures. As an integrated property, it shows smoother pattern with less deviation among databases as compared to the standard-state specific heat capacity at constant pressure.

As a quantitative measure of the deviation among databases in terms of the estimated thermodynamic properties, we present the maximum difference encountered between any two databases for each of the four standard-state properties for each of the six species. The values involved are those evaluated at the fourteen temperature values. We also include the temperature at which this maximum deviation occurs. These data are given in Table 5 for $O_2$, in Table 6 for $N_2$, in Table 7 for $H_2$, in Table 8 for $CH_4$, in Table 9 for $CO_2$, and in Table 10 for $H_2O$. Out of the 24 reported maximum deviations, 14 are located at the peak examined temperature point (3500 K). The other 10 maximum deviations are distributed over more than one temperature points.

## Conclusions

This work can be of interest to individuals or organizations working in computational modeling for problems with fluids, especially when high-temperatures, mixing, or reactions are involved.

Figure 7. Standard-state absolute engineering specific enthalpy of molecular oxygen and molecular nitrogen versus temperature, calculated using three different databases.

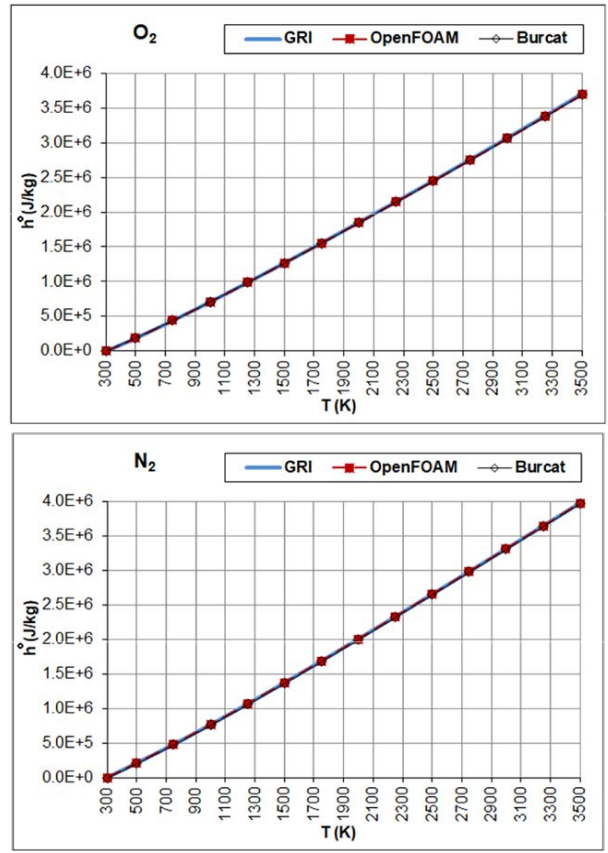

Figure 8. Standard-state absolute engineering specific enthalpy of molecular hydrogen and methane versus temperature, calculated using three different databases.

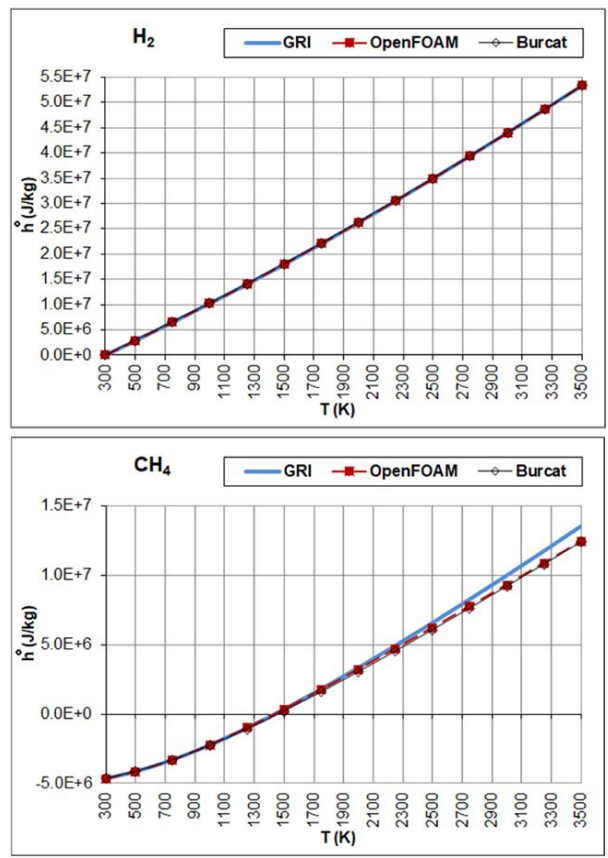

Figure 9. Standard-state absolute engineering specific enthalpy of carbon dioxide and water vapor versus temperature, calculated using three different databases.

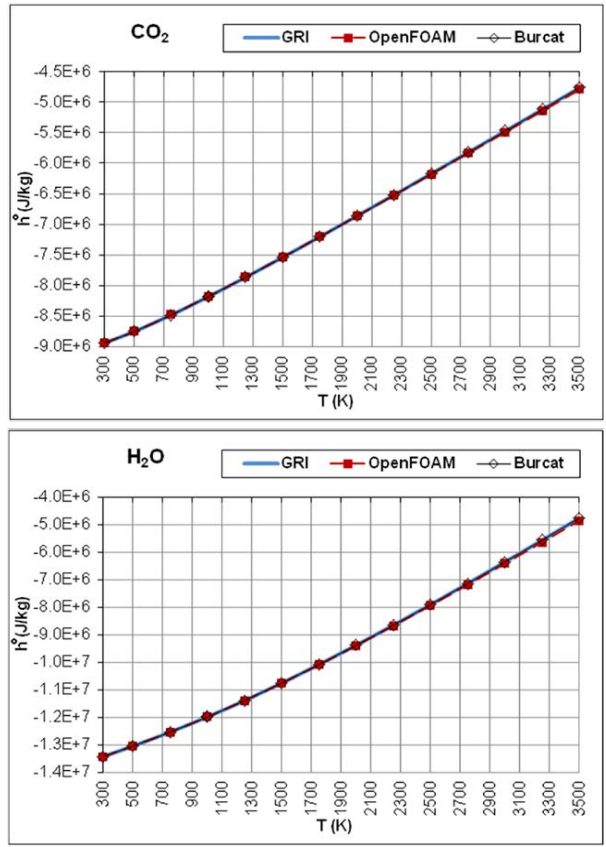

Figure 10. Standard-state mass-based specific entropy of molecular oxygen and molecular nitrogen versus temperature, calculated using three different databases.

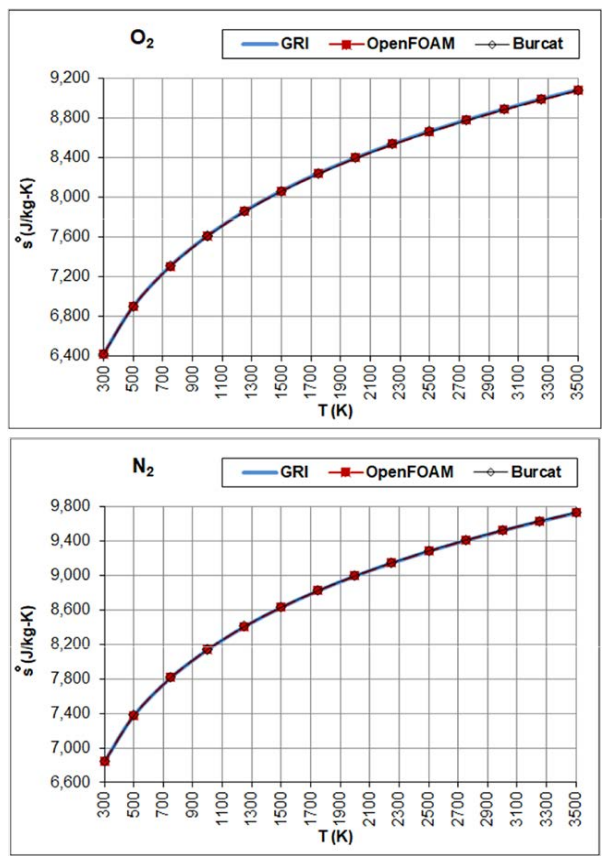

Figure 11. Standard-state mass-based specific entropy of molecular hydrogen and methane versus temperature, calculated using three different databases.

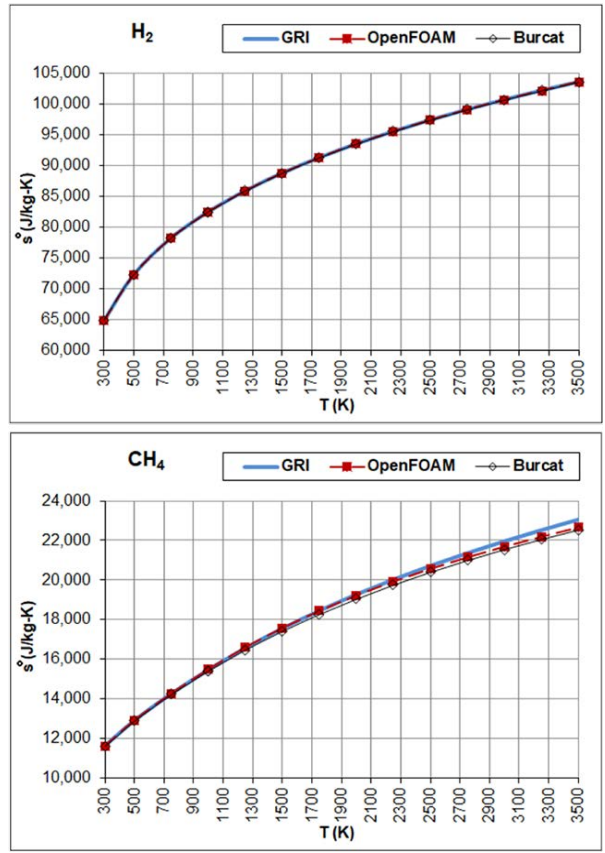

Figure 12. Standard-state mass-based specific entropy of carbon dioxide and water vapor versus temperature, calculated using three different databases.

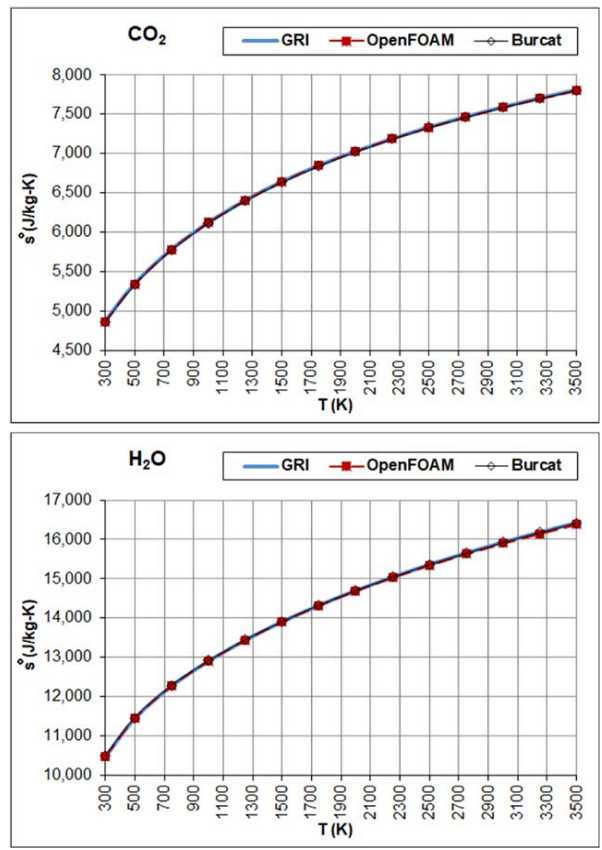

It presents the three NASA polynomial functions and their relation to four thermodynamic properties at a standard pressure. These properties are the specific heat capacity at constant pressure, specific heat ratio, specific enthalpy, and specific entropy. The polynomials include 7 coefficients per temperature interval per species.

The work then presented three accessible databases for the coefficients in the polynomials and analyzed them from a number of aspects to help deciding on the preferred one. The analysis utilized 6 common gaseous species ($O_2$, $N_2$, $H_2$, $CH_4$, $CO_2$, and $H_2O$) and 14 temperature points in the range of 300-3500 K.

The databases are the one accompanying the latest edition of the GRI-MECH reaction mechanism, the most-comprehensive one among those accompanying the tutorials of the latest version of OpenFOAM solver, and the latest edition of the Burcat thermodynamic data.

Overall, we recommend the Burcat database. Also, the specific heat capacity or specific heat ratio figures presented here can be used as class resources to emphasize the incurred errors of as-suming a constant value or ignoring variations across gaseous species in combustion or shock waves or other similar problems.

Table 5. Maximum encountered deviation in the estimated standard-state properties of ($O_2$) when estimated at 14 temperatures between 300 K and 3500 K. The temperature corresponding to each maximum deviation is also given.

| Property | $c_p^\circ \left( \dfrac{J}{kg.K} \right)$ | $h^\circ \left( \dfrac{J}{kg} \right)$ | $\gamma^\circ$ | $s^\circ \left( \dfrac{J}{kg.K} \right)$ |
|---|---|---|---|---|
| Maximum deviation | 6.5 | 5982 | 1.7E-03 | 4.9 |
| Temperature (K) | 3250 | 3500 | 3250 | 3500 |

Table 6. Maximum encountered deviation in the estimated standard-state properties of ($N_2$) when estimated at 14 temperatures between 300 K and 3500 K. The temperature corresponding to each maximum deviation is also given.

| Property | $c_p^\circ \left( \dfrac{J}{kg.K} \right)$ | $h^\circ \left( \dfrac{J}{kg} \right)$ | $\gamma^\circ$ | $s^\circ \left( \dfrac{J}{kg.K} \right)$ |
|---|---|---|---|---|
| Maximum deviation | 2.8 | 1525 | 1.1E-03 | 3.5 |
| Temperature (K) | 1000 | 1750 | 1000 | 300 |

Table 7. Maximum encountered deviation in the estimated standard-state properties of ($H_2$) when estimated at 14 temperatures between 300 K and 3500 K. The temperature corresponding to each maximum deviation is also given.

| Property | $c_p^\circ \left( \dfrac{J}{kg.K} \right)$ | $h^\circ \left( \dfrac{J}{kg} \right)$ | $\gamma^\circ$ | $s^\circ \left( \dfrac{J}{kg.K} \right)$ |
|---|---|---|---|---|
| Maximum deviation | 56.3 | 38090 | 1.5E-03 | 68.4 |
| Temperature (K) | 2500 | 3500 | 1250 | 500 |

Table 8. Maximum encountered deviation in the estimated standard-state properties of ($CH_4$) when estimated at 14 temperatures between 300 K and 3500 K. The temperature corresponding to each maximum deviation is also given.

| Property | $c_p^\circ \left( \dfrac{J}{kg.K} \right)$ | $h^\circ \left( \dfrac{J}{kg} \right)$ | $\gamma^\circ$ | $s^\circ \left( \dfrac{J}{kg.K} \right)$ |
|---|---|---|---|---|
| Maximum deviation | 918.0 | 1118689 | 1.2E | 513.9 |
| Temperature (K) | 3500 | 3500 | 3500 | 3500 |

Table 9. Maximum encountered deviation in the estimated standard-state properties of ($CO_2$) when estimated at 14 temperatures between 300 K and 3500 K. The temperature corresponding to each maximum deviation is also given.

| Property | $c_p^\circ \left( \dfrac{J}{kg.K} \right)$ | $h^\circ \left( \dfrac{J}{kg} \right)$ | $\gamma^\circ$ | $s^\circ \left( \dfrac{J}{kg.K} \right)$ |
|---|---|---|---|---|
| Maximum deviation | 27.3 | 32807 | 5.2E | 9.3 |
| Temperature (K) | 3500 | 3500 | 300 | 3500 |

**Table 10.** Maximum encountered deviation in the estimated standard-state properties of ($H_2O$) when estimated at 14 temperatures between 300 K and 3500 K. The temperature corresponding to each maximum deviation is also given.

| Property | $c_p^\circ \left(\dfrac{J}{kg.K}\right)$ | $h^\circ \left(\dfrac{J}{kg}\right)$ | $\gamma^\circ$ | $s^\circ \left(\dfrac{J}{kg.K}\right)$ |
|---|---|---|---|---|
| Maximum deviation | 68.3 | 92315 | 4.2E | 41.3 |
| Temperature (K) | 3500 | 3500 | 3500 | 3500 |